\documentclass{WileyASNA-v1}

\articletype{ORIGINAL ARTICLE}

\received{}
\revised{}
\accepted{}

\usepackage{hyperref}
\usepackage{graphics}

\usepackage{xspace}
\usepackage{url}

\usepackage{natbib}

\def\swift{\textit{Swift}\xspace}
\def\nus{\textit{NuSTAR}\xspace}
\def\xmm{\textit{XMM-Newton}\xspace}
\def\cha{\textit{Chandra}\xspace}
\def\suz{\textit{Suzaku}\xspace}
\def\fermi{\textit{Fermi}\xspace}
\def\hessj{HESS~J0632+057\xspace}

\def\psrbl{PSR~B1259-63/LS2883\xspace}
\def\nh{$n_\mathrm{H}$\xspace}

\begin{document}
	%\linenumbers 
	\title{Decade-long X-ray observations of HESS J0632+057} 
	\author[1]{D. Malyshev}
	\author[2,3]{M. Chernyakova}
	\author[1]{A. Santangelo}
	\author[1]{G. P\"uhlhofer}
	
	\authormark{D.~Malyshev~\textsc{et al.}}
	
    \address[1]{Institut f{\"u}r Astronomie und Astrophysik T{\"u}bingen, Universit{\"a}t T{\"u}bingen, Sand 1, D-72076 T{\"u}bingen, Germany}
    \address[2]{School of Physical Sciences and C-fAR, Dublin City University, Dublin 9, Ireland}
    \address[3]{Dublin Institute for Advanced Studies, 31 Fitzwilliam Place, Dublin 2, Ireland
    }
    
\abstract{We present results of a decade of X-ray observations of the gamma-ray loud binary \hessj, and interpret the available broadband data in view of the system geometry and emission mechanisms. We have performed an analysis of all X-ray data available to date from \swift, \xmm, \cha, \nus and \suz. We refine the orbital period of the system to be $316.8^{+2.6}_{-1.4}$~d (95\% c.l.), consistent with previous studies but measured with significantly better accuracy. We report on a hydrogen column density and spectral slope variation along the orbit. We argue that the observed variability can be explained within an ``inclined disk'' model in which the orbit of the compact object is inclined to the disk of the Be star. We show that the observed X-ray to TeV emission can originate from a broken cut-off power-law population of electrons and describe a way in which future X-ray/TeV observations can distinguish between the proposed model and the alternative flip-flop emission scenario of this system.}
%\date{Received $<$date$>$  ; in original form  $<$date$>$ }
%\pagerange{\pageref{firstpage}--\pageref{lastpage}} \pubyear{2017}

\jnlcitation{\cname{%
\author{D.~Malyshev}, 
\author{M. Chernyakova}, 
\author{A. Santangelo},
\author{G.P\"uhlhofer}
} (\cyear{2019}), 
\ctitle{Decade-long X-ray observations of HESS J0632+057}, 
\cjournal{ASNA}, \cvol{}.}

%\fundingInfo{Funding info text.}

\maketitle

\section{Introduction} 
\label{sec:intro} 
\hessj belongs to the rare class of gamma-ray loud binaries (GRLBs). These are peculiar systems with a spectral energy distribution (SED) dominated by emission in the GeV--TeV band. All known systems of this class host a compact object that is orbiting either an O- or a Be-type star. These types of stars are characterized by strong stellar outflows (circumstellar disks in case of Be stars).

The SEDs similarity of gamma-ray loud binaries suggests that the compact objects are of similar nature and that similar physical mechanisms are at work in the systems. Differences are mainly attributed to the geometry of the system and to the viewing angle.
At least in two GRLBs, \psrbl, and PSR J2032+4127/MT91~213, the compact objects are known to be pulsars \citep{Johnston1992,psrj2032fermi,lyne15}, which suggests that neutron stars are the compact objects in other GRLBs as well~\citep{dubus06,neronov08,torres10,zdz10,durant11,moritani15,bird17}. However, a black hole nature of the compact object has been suggested for other GRLBs (see e.g., \citealt{massi01,casares05,williams10,massi17} and references therein).
%%%%%%%%%%%%%%%%%%%%%%%%%%%%%%%%%%%%%
\begin{table*}
\begin{tabular}{|c|c|c|c|c|c|c|}
\hline
Label&Obs Id & Date, MJD / & \nh, $10^{22}$~cm$^{-2}$ & log$_{10}$($F_{0.3-10}$/1erg/cm$^2$/s)  & $\Gamma_{0.3-10}$ & Comment \\
& & Orbital phase& &  &  &  \\
\hline
X1 &0505200101 & 54360 / -1.57 & 0.31 $\pm$ 0.05 & -12.29 $\pm$ 0.02 & 1.26 $\pm$ 0.08 & MOS~1+2 \\
X2 & 0821370201 & 58371 / 11.09  & 0.28 $\pm$ 0.03 & -11.85 $\pm$ 0.01 & 1.6 $\pm$ 0.05 & \\
S1 &403018010  & 54579 / -0.88  & 0.27 $\pm$ 0.025& -11.76 $\pm$ 0.01 & 1.55 $\pm$ 0.035&  XIS~0+1+3 \\
S2 &404027010  & 54941 / 0.27 & 0.22 $\pm$ 0.016& -11.768 $\pm$ 0.005&1.37 $\pm$ 0.02 &  XIS~0+1+3\\
C1 &13237      & 55605 / 2.36  & 0.47 $\pm$ 0.025& -11.46 $\pm$ 0.01 & 1.67 $\pm$ 0.045&  asic-cc\\
C2 & 20269     & 58152 / 10.40 & 0.11 $\pm$ 0.14 & -12.23 $\pm$ 0.04 & 1.31 $\pm$ 0.17 &  \\
C3 & 20950     & 58153 / 10.40  & 0.30 $\pm$ 0.16 & -12.33 $\pm$ 0.04 & 1.45 $\pm$ 0.18 &  \\ 
C2+C3 & --     & --    & 0.22 $\pm$ 0.11 & -12.29 $\pm$ 0.03 & 1.42 $\pm$ 0.13 & \\
N1 &30362001002& 58079 / 10.17  & 0.3*            & (-11.76 $\pm$ 0.01)* & (1.72 $\pm$ 0.04)* &  A+B \\
N2 &30362001004& 58101 / 10.24  & 0.3*            & (-11.80 $\pm$ 0.01)* & (1.59 $\pm$ 0.04)* &  A+B\\
N3 & 30401006002 & 58371 / 11.09 & 0.28* & (-11.88 $\pm$ 0.02)* & (1.72 $\pm$ 0.06)* &  A+B\\
\hline
\end{tabular}
\caption{Summary of the analyzed \xmm (X1-X2), \cha(C1-C3), \suz(S1-S2) and \nus(N1-N3) observations. The data were fitted in the $0.3-10$~keV range with an absorbed power-law model (\texttt{cflux*phabs*po}) with a hydrogen column density \nh, a slope $\Gamma_{0.3-10}$, and a flux in 0.3-10~keV range $F_{0.3-10}$, except for the \nus observations (results marked with *) which were fitted in the energy range $2-50$~keV, the hydrogen column density was fixed to 0.3, and the best-fit flux values were rescaled to the 0.3-10~keV band.}
\label{tab:xray_obs}
\end{table*}
%%%%%%%%%%%%%%%%%%%%%%%%%%%%%%%%%%%%%

\hessj was discovered as a serendipitous source in the field of H.E.S.S.\ observations of the Monoceros region \citep{hess_j0632}. The spatial coincidence with the Be star MWC~148 \citep{hess_j0632} as well as the properties of the soft X-ray and radio counterparts \citep{hinton09,skilton09,falcone10} strongly suggested a galactic binary nature of the source.
The peculiar position of \hessj also allowed its detection at TeV energies by MAGIC and VERITAS~\citep{magic12,veritas13} observatories operating in the northern hemisphere.
Contrary to other known GRLBs, \hessj has not been firmly detected at GeV energies, most likely because it lies in a particularly crowded the region in the GeV band. Only recently hints of a detection with \fermi/LAT were reported by~\citet{we_hessj,li17}.

\swift-XRT observations of \hessj performed between 2009 and 2011 helped to determine the period of the system ($P_\mathrm{orb}= 320 \pm 5$~d, $T_0$=MJD~54857.0, see e.g. \citealt{bongiorno11}). This period was then used to determine the other orbital parameters from optical observations: eccentricity $e=0.83 \pm 0.08$, periastron phase $\phi_0=0.967 \pm 0.008$, the argument of periastron $\omega=129\pm 17$, inclination $i\gtrsim 47^\circ$, and distance in apastron $d_\mathrm{apastron}\sim 5$~a.u. (\citealt{casares12}). The distance to the source was estimated to be $\sim 1.4$~kpc~\citep{casares12}\footnote{Note, that a much higher distance $2.8\pm 0.3$~kpc was reported in GAIA data release 2~\citep{gaia_dr2} }. The orbital period was later refined by \citet{veritas13} ($P_\mathrm{orb}=315^{+6}_{-4}$~d) using additional \swift-XRT observations of 2012.
Two clear X-ray emission peaks are observed in the orbital cycle of \hessj, the first at phase $\phi\sim 0.2-0.4$ and the second at $\phi\sim 0.6-0.8$, separated by a local flux minimum at $\phi\sim 0.4-0.5$ \citep{bongiorno11,veritas13}. A similar structure of the TeV orbital lightcurve has been reported by \citet{veritas15}. Hints of orbital variability in the GeV range were reported in \citet{li17}.

Two are the main goals of this work: 1) to perform a consistent analysis of all available X-ray data \footnote{Observations from \swift, \xmm, \cha, \nus and \suz are available}, including previously unpublished 2017--2018 \swift ToO observations of the first emission peak to study the orbital resolved X-ray emission and the source's spectral energy distribution (SED); and 2) to discuss the observed phenomenology in the framework of an ``inclined disk'' model in which the orbit of the compact object is inclined with respect to the disk of the Be star, and the emission arises from relativistic electrons accelerated in the pulsar wind shock at all orbital phases. 

In section 2, we present the details of the data analysis. In section 3, an improved value of the orbital period is derived, and we report about a significant variability of the hydrogen column density \nh, and of the X-ray spectral slope along the compact object's orbit. In section 4, the X-ray orbital data and the SED are confronted with the proposed model, and we present observational signatures that could be used in the future to discriminate between our proposed model and the alternative flip-flop model.

\section{Data Analysis}
\label{sec:data_analysis}
All available X-ray data of \hessj were analyzed using the most recent calibration files and data analysis software. Data reduction was performed with the most recent available \texttt{heasoft v.6.22} software package, while spectral analysis was performed with \texttt{XSPEC v.12.9.1m}. All spectra were grouped to a minimum of 1~count/bin and were fitted in the 0.3-10~keV energy range with an absorbed power law model (\texttt{cflux*phabs*po}), using the c-statistics suitable for the analysis of poor-statistics data\footnote{See \href{https://heasarc.gsfc.nasa.gov/xanadu/xspec/manual/XSappendixStatistics.html}{description of statistics used in XSPEC}}).
The best-fit values for the flux, spectral slope, and hydrogen column density are summarized in Table~\ref{tab:xray_obs}, and are shown as a function of orbital phase bin in Figs.~\ref{fig:orbital_profile} and \ref{fig:orbital_nh_idx}. Instrument-specific details of the data analysis are briefly summarized below. Within statistical uncertainties, our results are consistent with previously published results where available (\citealt{bongiorno11,veritas13,veritas15} (\swift-XRT, partially),\\ \citealt{rea11} (\xmm and \cha), \citealt{hinton09,bongiorno11} (\xmm), \citealt{skilton11} (\suz-XIS)). 

\subsection*{\swift-XRT data analysis}
Publicly available \swift-XRT data on \hessj have been taken between January 26th, 2009 and February 20th, 2018. The data were reprocessed with \texttt{xrtpipeline v.0.13.4} as suggested by the \swift-XRT team\footnote{See e.g.\ the \href{https://swift.gsfc.nasa.gov/analysis/xrt_swguide_v1_2.pdf}{\swift-XRT User's Guide}}.
Spectra were extracted with \texttt{xselect} using the coordinates of the optical counterpart\footnote{(RA;Dec) = (98.246894 ; 5.800327) from~\citep{coords}} of \hessj , using a $36''$ circle for source counts and an annulus also centered at the source position with inner (outer) radii of $60''$ ($300''$) for background counts. 

\subsection*{\xmm data analysis}
The analysis of the only available \xmm observation of \hessj (taken on Sept. 17, 2007 and labeled ``X1'' hereafter) was performed with the latest \xmm \texttt{Science Analysis software v.15.0.0}. Known hot pixels and electronic noise were removed, and data were filtered to exclude soft proton flares episodes. The total exposure is $\sim 30$~ksec. Unfortunately, since \hessj was located close to the border of one PN camera chip, the corresponding data were affected by soft proton flares. For a conservative analysis, we therefore only used data from the MOS cameras. The spectrum was extracted from a $40''$ radius circle centered at the position of \hessj and the background was extracted from a nearby source-free region of $80''$ radius. The RMFs and ARFs were extracted using the \texttt{RMFGEN} and \texttt{ARFGEN} tools, respectively. 

\subsection*{\cha data analysis}
\begin{figure*}
\includegraphics[width=0.47\linewidth]{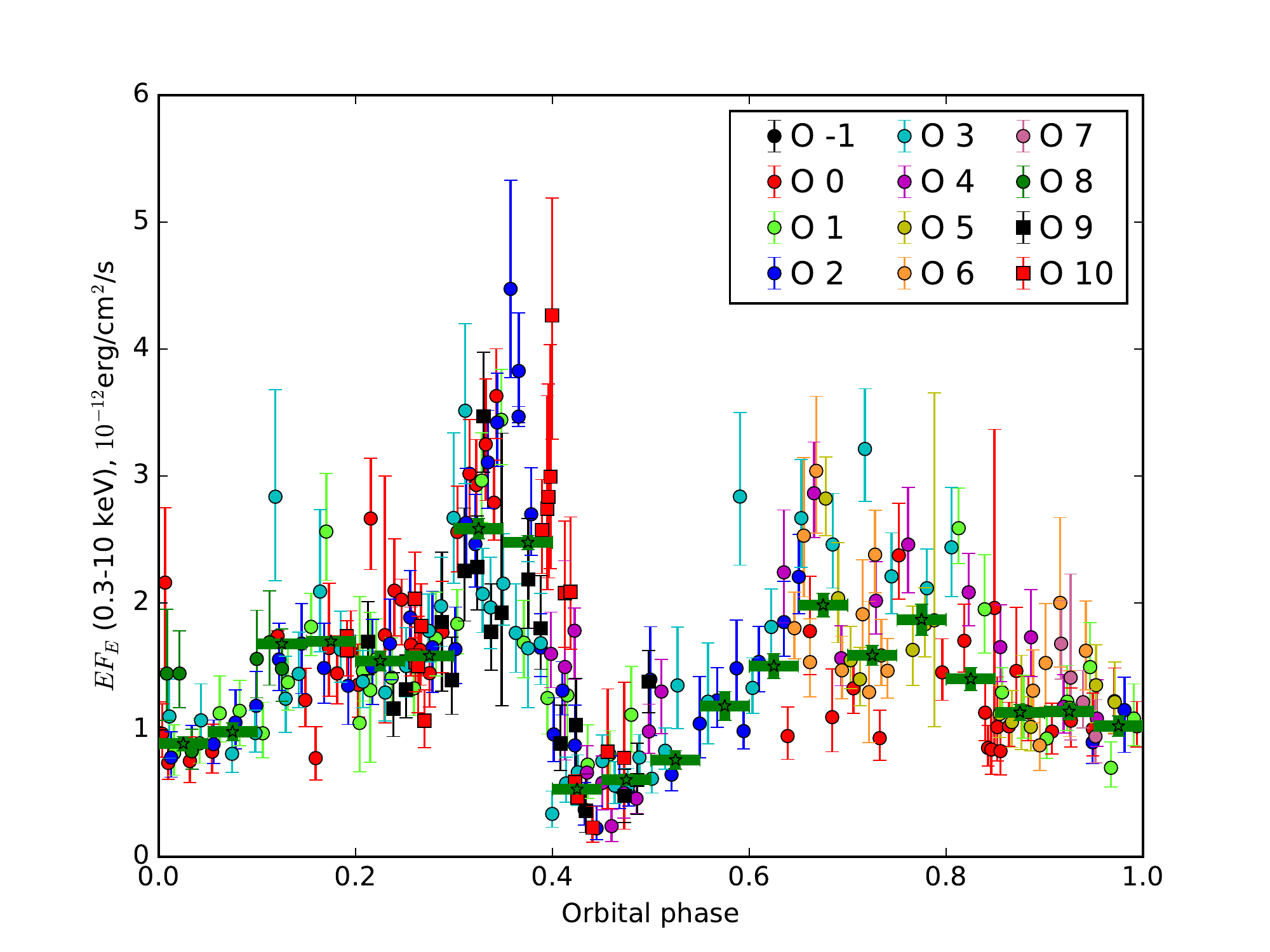}
\includegraphics[width=0.47\linewidth]{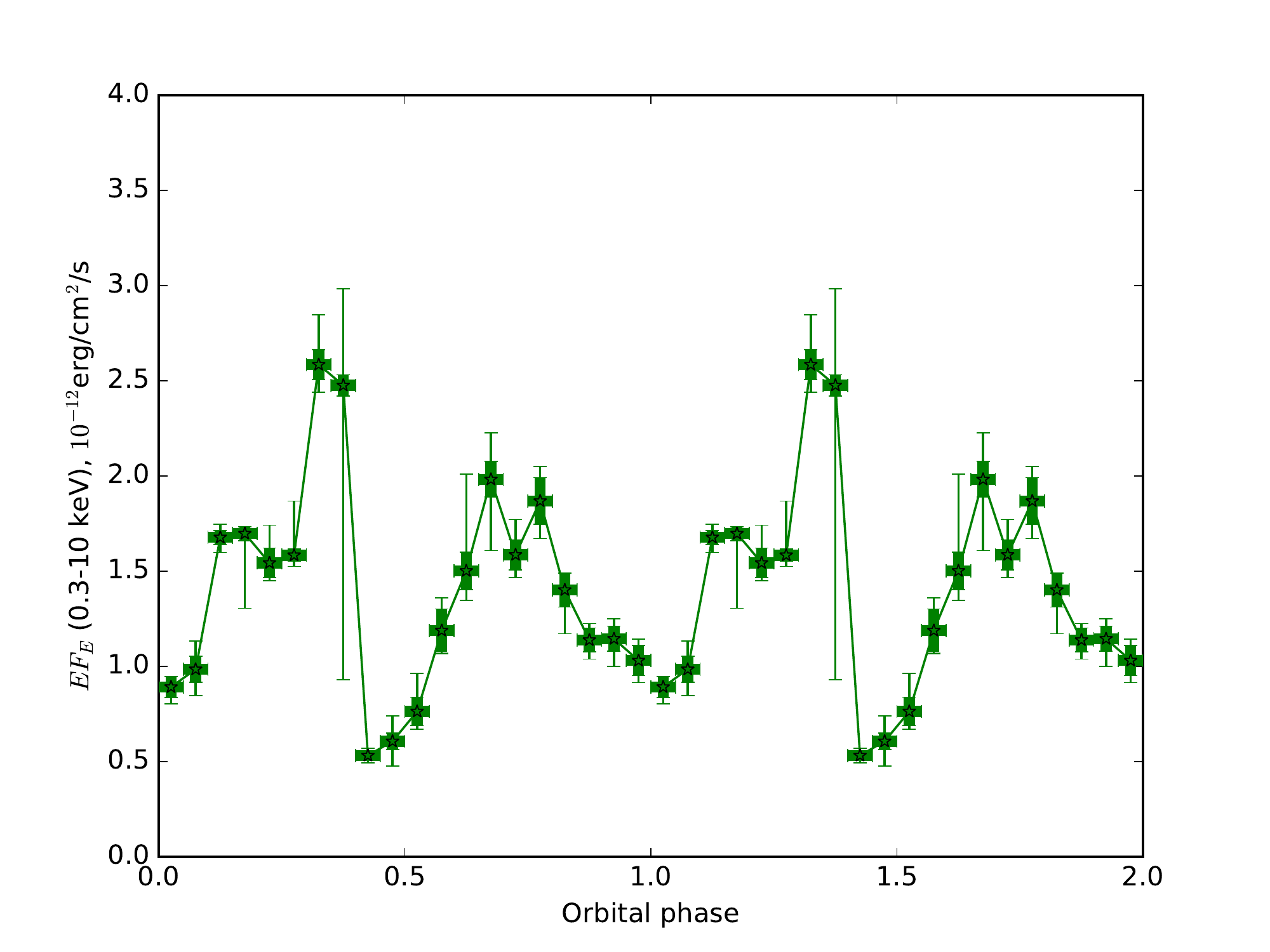}
\caption{The flux orbital profile of \hessj for the best-fit period $P_\mathrm{orb}=316.8$~d. \textit{Left:} Orbital profiles for individual orbits ($N$-th orbit data is marked with ``O $N$'' ; green points with the star markers show the fluxes averaged over all orbits). \textit{Right:} Average flux orbital profile from a mean over all orbits. Thick error bars show the statistical uncertainties for the best-fit orbital period, while the thin ones illustrate the uncertainty of the profile from the uncertainties of the period determination (95\% c.l.).}
\label{fig:orbital_profile}
\end{figure*}

\begin{figure*}
\includegraphics[width=0.47\linewidth]{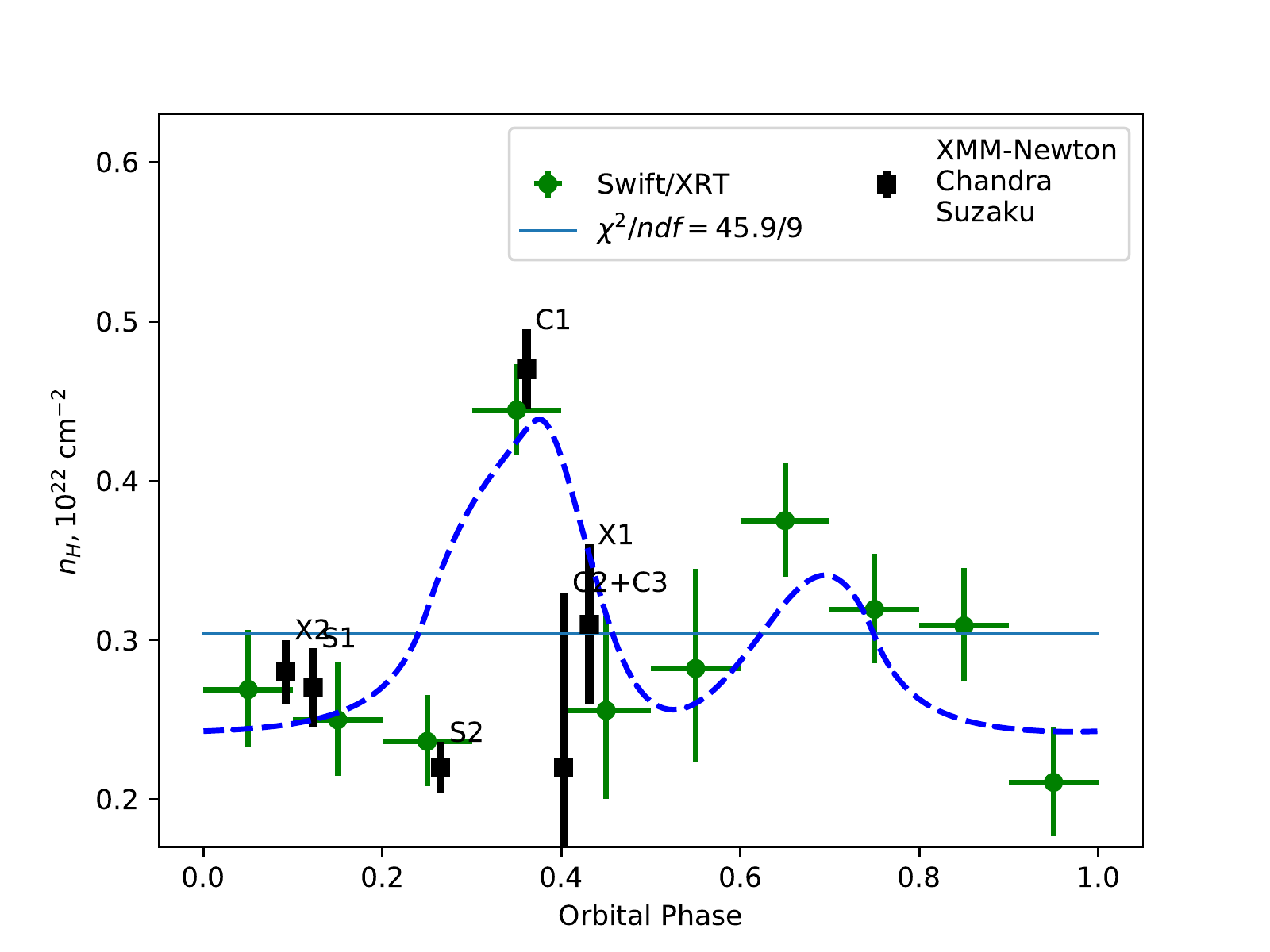}
\includegraphics[width=0.47\linewidth]{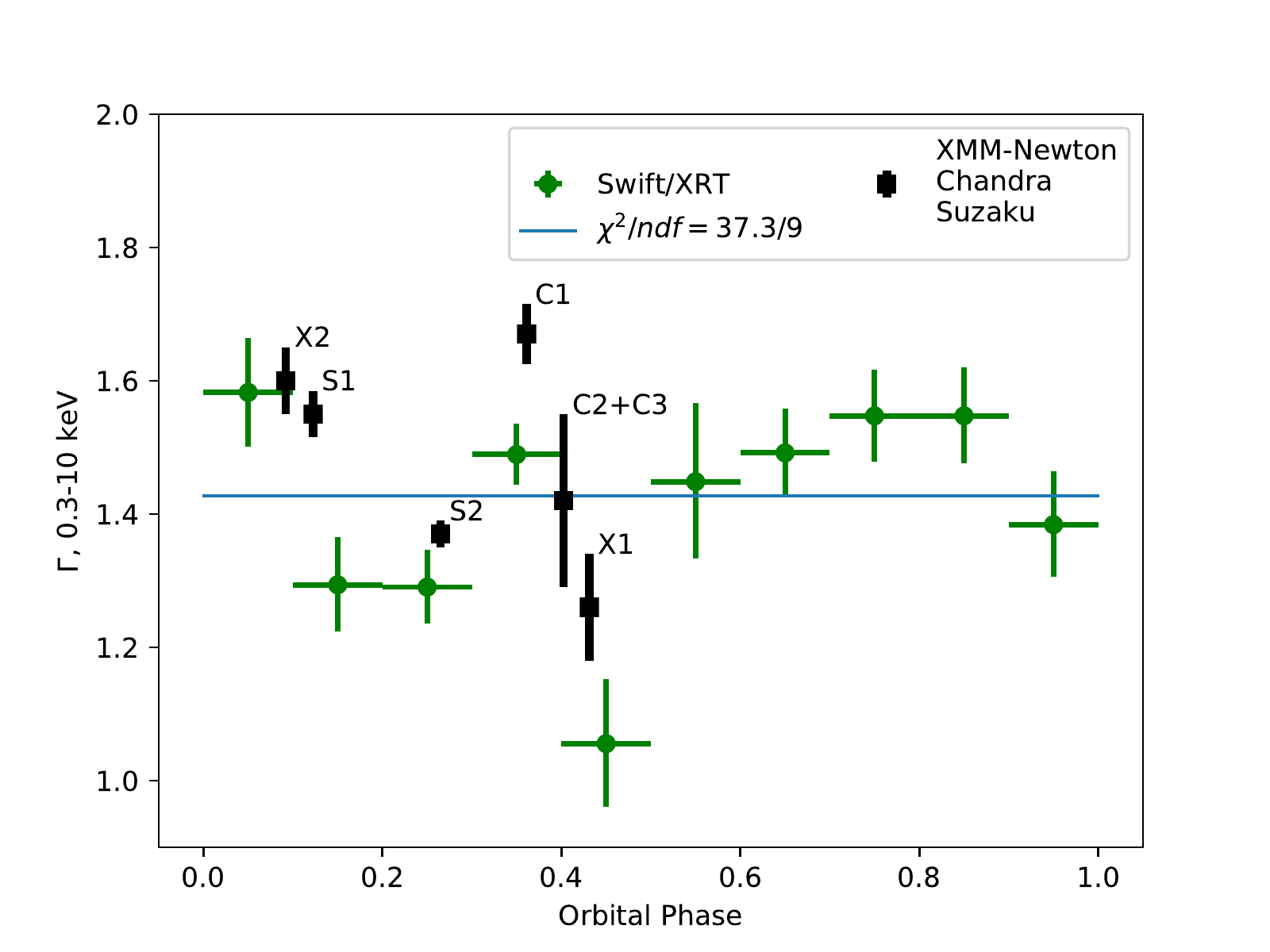}
\caption{Orbital profiles of the hydrogen column density \nh (left panel) and the $0.3-10$~keV spectral slope $\Gamma$ of \hessj. Black points show the high-quality results from individual observations of \xmm, \cha, and \suz. Thin blue lines show the best fit with a constant, with $\chi^2$-values as given in the insets (not taking the black points into account). For the \nh-panel, the dashed blue line illustrates an \nh profile expected from a simple geometrical model of an edge-on oriented binary system with inclined disk, see text for details.}
\label{fig:orbital_nh_idx}
\end{figure*}

\cha observations of \hessj were performed in a high state in 2011, Feb. 13 and in a low flux state in Feb. 2--3, 2018 (see Figs.~\ref{fig:orbital_profile},\ref{fig:orbital_nh_idx}). To avoid confusion with the terminology accepted for accreting X-ray binary systems we would like to stress that hereafter, ``high-'' and ``low-'' states of the system refer to the corresponding orbital phase intervals for which high/low X-ray fluxes are consistently observed. We analyzed the data using the most recent \texttt{CIAO v.4.9} software and CALDB 4.7.6. The data was reprocessed with the \texttt{chandra\_repro} utility, source and background spectra with corresponding RMFs and ARFs were extracted with the \texttt{specextract} tool. The high-state observation (labeled ``C1'') was performed in asic-cc (continuous clocking) mode, in which only 1-dimensional spatial information is available. To extract in this case source and background spectra, we used box-shaped regions\footnote{See \href{http://cxc.harvard.edu/ciao/caveats/acis_cc_mode.html}{caveats of asic-cc mode data analysis}} centered on \hessj and located in a nearby source-free region, respectively. For the low-state observations (labeled ``C2'' and ``C3'') we utilized standard circle-shaped regions. The total exposure for observation C1 was $\sim 40$~ksec, for observations C2 and C3 33~ksec and 45~ksec, respectively. Given the high uncertainties for the best-fit spectral parameters during the low-state \cha observations, we also combined the C2 and C3 datasets, see Table~\ref{tab:xray_obs}.

%%%%%%%%%%%%%%%%%%%%%%%%%%%%%%%%%%%%%
\begin{figure}
\includegraphics[width=1.05\linewidth]{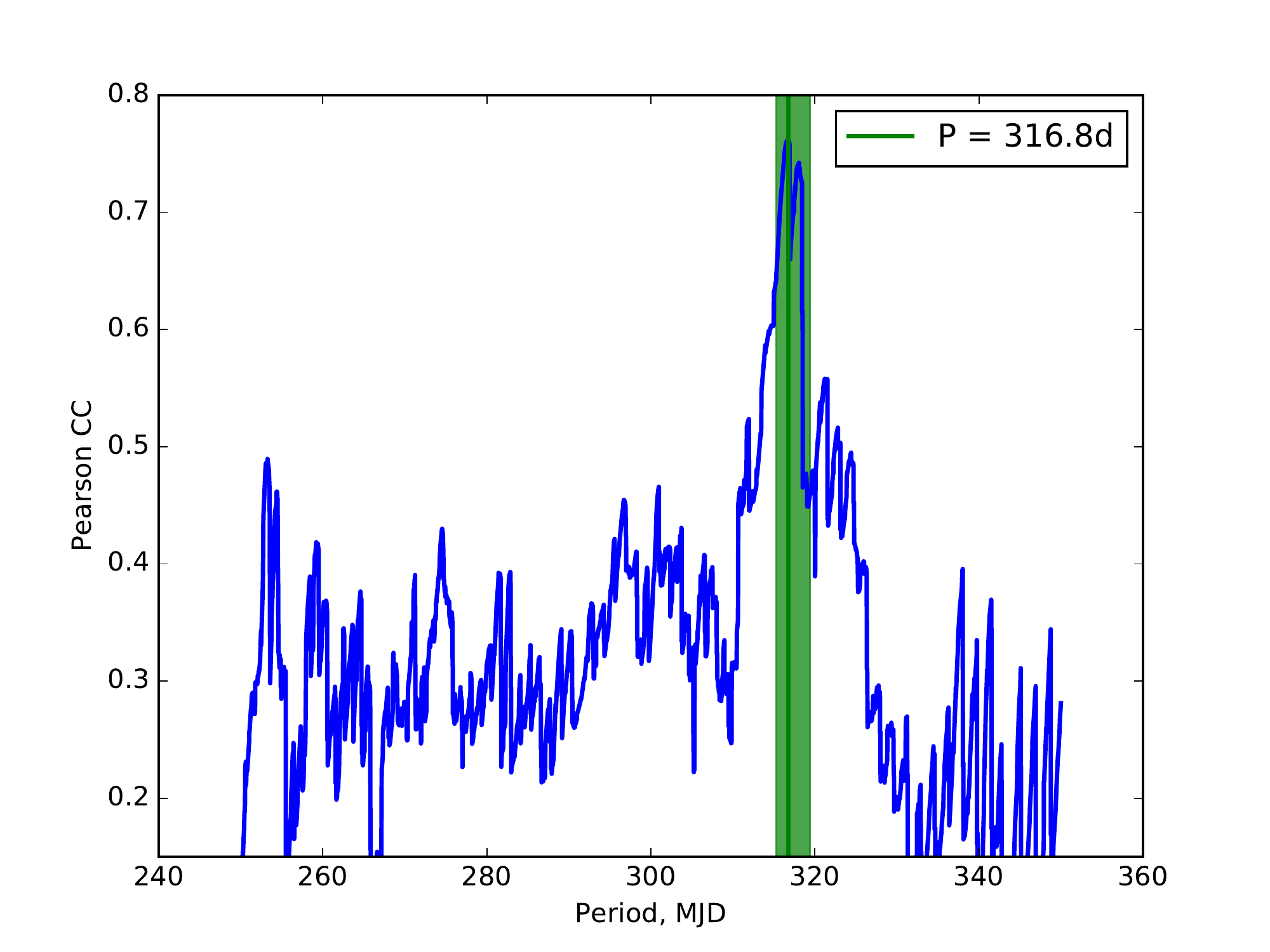}
\caption{The Pearson correlation coefficient derived from the relation between mean observed fluxes per phase bin in one orbit and model fluxes, for a range of assumed orbital periods. Model fluxes are derived from a smoothed light curve average of the data in one phase bin over all observed orbits, see Sec.~\ref{sec:period_search} for details. The vertical green line indicates the maximum of the Pearson correlation coefficient, observed at $P_{orb}=316.8$~d. The shaded area illustrates the 95\% confidence region (statistical and systematic uncertainties) for the orbital period value.}
\label{fig:autocorrelation}
\end{figure}
%%%%%%%%%%%%%%%%%%%%%%%%%%%%%%%%%%%%%

\subsection*{\suz data analysis}
Two \suz observations were performed on \hessj, on Apr. 23, 2008 and Apr. 20, 2009, labeled ``S1'' and ``S2'' in the following. For the analysis we used data of XIS 0, 1, and 3, reprocessed with the \texttt{aepipeline v.1.1.0} tool. The source spectrum as well as the corresponding response files were extracted with the \texttt{xselect} tool\footnote{as described in the \href{https://heasarc.gsfc.nasa.gov/docs/suzaku/analysis/abc/node9.html}{Suzaku ABC Guide}} from a circle centered at \hessj and with a radius of $150''$. 
The background spectrum was extracted from an annulus centered at the same position and inner/outer radii of $151''$ and $288''$. We also find that the source is not detected with \suz-HXD/PIN, with upper limits consistent with a power-law extrapolation of the XIS best-fit data model.
%%%%%%%%%%%%%%%%%%%%%%%%%%%%%%%%%%%%%
%%%%%%%%%%%%%%%%%%%%%%%%%%%%%%%%%%%%%

%%%%%%%%%%%%%%%%%%%%%%%%%%%%%%%%%%%%%
%%%%%%%%%%%%%%%%%%%%%%%%%%%%%%%%%%%%%
\subsection*{\nus data analysis}
Two \nus observations were performed on \hessj, on Nov. 22nd and Dec. 14th, 2017, labeled ``N1'' and ``N2'', see Table~\ref{tab:xray_obs}. The raw data were processed with standard pipeline processing (HEASOFT v.6.22 with the NuSTAR subpackage v.1.8.0). We applied strict criteria for the exclusion of data taken in the South Atlantic Anomaly (SAA) and in the ``tentacle''-like region of higher background activity near part of the SAA. Level-two data products were produced with the \texttt{nupipeline} tool with the flags \texttt{SAAMODE=STRICT} and \texttt{TENTACLE=yes}. High-level spectral products (spectra, response matrices, and auxiliary response files) were extracted for a point source with the \texttt{nuproducts} routine. The corresponding background flux was derived from a ring-like (inner/outer radii of $80''$/$196.8''$) region surrounding the source. The spectral analysis was performed in the energy range of $2-50$~keV. Since the energy band $\lesssim 2$~keV is not available from \nus, the hydrogen column density cannot be robustly constrained. For the spectral analysis we thus adopt a value for $n_\mathrm{H}$ fixed to $0.3\cdot 10^{22}$~cm$^{-2}$.

\section{Results}
\label{sec:results}

\subsection{Orbital periodicity search}
\label{sec:period_search}
The data described in Sect.~\ref{sec:data_analysis}, including the most recent \swift-XRT ToO observations taken during the high-flux orbital state of the source, spans over about 10 orbits of \hessj. This allowed us to improve the measurement of the orbital period compared to previous works \citep{bongiorno11,veritas13}. 

We performed an autocorrelation analysis for assumed periods ranging from 250~d to 350~d, with a step size of 0.1~d. For each of these periods, as the first analysis step we binned the data in 20 orbital phase bins of equal duration, defining the flux in each bin as the weighted (with corresponding flux uncertainties) mean of all fluxes whose observations are attributed to this phase bin.
At this point we explicitly assume that the orbital profile is smooth on scales of $\sim 0.05$~orbit and that the average number of observations per selected bin is large enough for the uncertainty on the mean profile to be significantly smaller than the uncertainty of each individual measurement.

Based on this binned orbital flux profile, for each assumed period we define a smoothed orbital flux profile from linear interpolations between fluxes in each of the neighboring bins (with the time of the flux in each bin taken as the center of the bin), with periodic boundary conditions. This smoothed orbital flux profile defines a model flux at the exact time of each individual observation. As a second step of the analysis we compute the Pearson correlation coefficient between model and observed fluxes as a function of the assumed orbital period, see Fig.~\ref{fig:autocorrelation}. The correlation coefficient exhibits a clear maximum at the orbital period $P_\mathrm{orb}=316.8$~d, shown in Fig.\,\ref{fig:autocorrelation} with a vertical solid green line.

In order to estimate the statistical uncertainty of $P_\mathrm{orb}$, $10^4$ random-trial realizations of the flux data points have been simulated based on the observed flux values and their corresponding uncertainties. For each realization $r$ of the data we repeated the analysis as described above and determined $P_\mathrm{orb}(r)$. We define the 95\% statistical confidence range for $P_\mathrm{orb}$ as a symmetrically-centered region containing 95\% of all $P_\mathrm{orb}(r)$ (0.025 -- 0.975 quantiles). 

We checked for systematic effects caused by the selected number of phase bins on the measured orbital period, by varying the number of bins from 10 to 40 and performing each time the random-trial analysis as described above. We define a 95\% confidence range accounting for this systematic effect by the interval which includes all statistical 95\% confidence intervals for $P_\mathrm{orb}$. 
The systematic uncertainty is comparable with the statistical error, the final result is
$P_\mathrm{orb}=316.8^{+1.2_\mathrm{stat}+1.4_\mathrm{syst}}_{-0.4_\mathrm{stat}-1.0_\mathrm{syst}}$~d. 
The error is shown as shaded green area in Fig.~\ref{fig:autocorrelation}.
%%%%%%%%%%%%%%%%%%%%%%%%%%%%%%%%%%%%%
\begin{figure}
\includegraphics[width=1.05\linewidth]{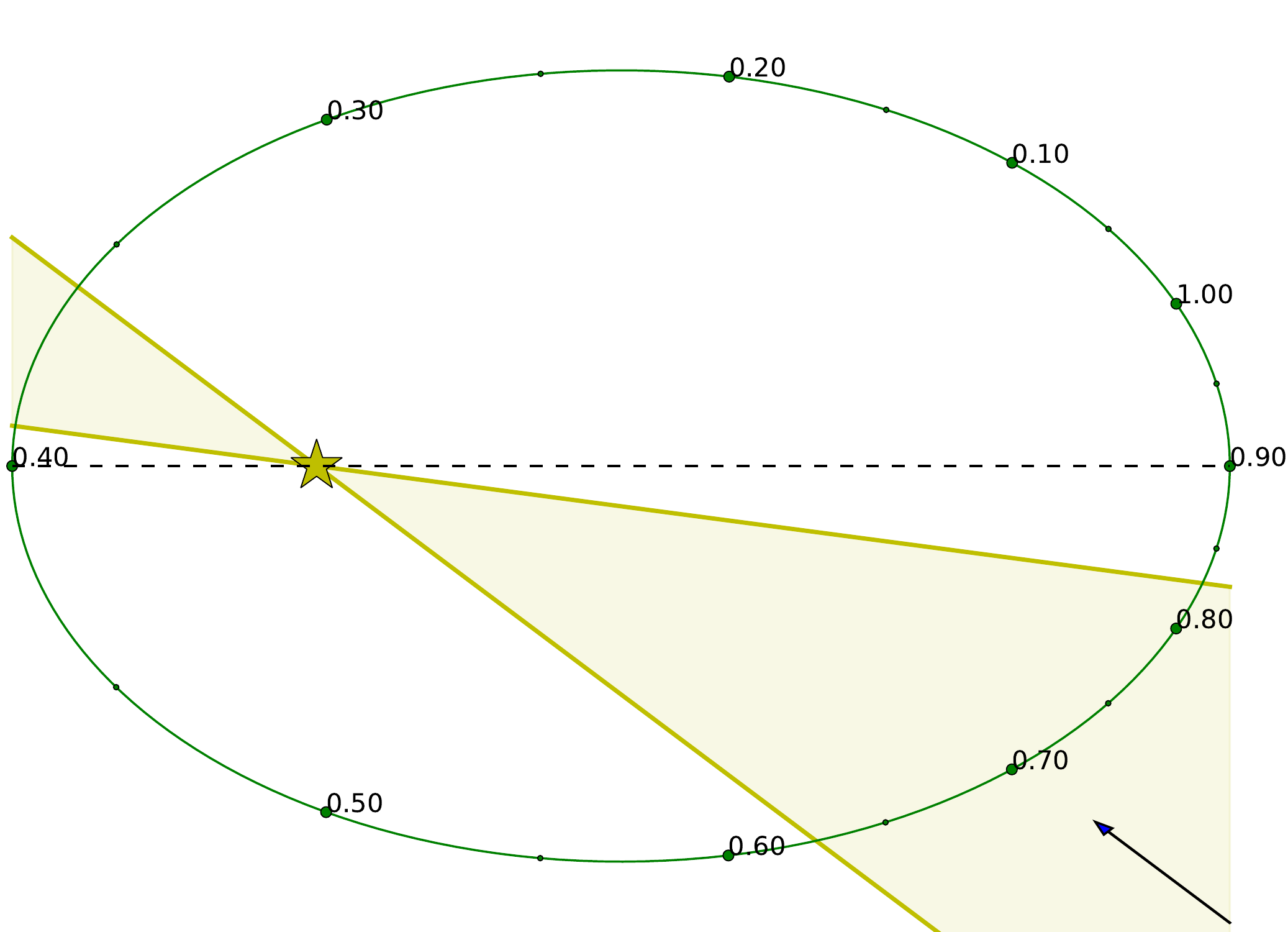}
\caption{Sketch of the \hessj system geometry viewed perpendicular to the orbital plane, for the orbital parameters preferred for the ``inclined disk'' model (eccentricity $\sim 0.5$, $\phi_{per}=0.4$). The disk of the Be star (illustrated with the region filled with light yellow; not to the scale, see Eq.~\ref{eq:disk} for the disk profile) is inclined to the orbital plane (solid green line). The green points along the orbit indicate the locations of every 0.05 orbital phase. The observer (black arrow) is located at a high inclination close to orbital plane.}
\label{fig:orbit}
\end{figure}
%%%%%%%%%%%%%%%%%%%%%%%%%%%%%%%%%%%%%

The observed fluxes, the observation times of which are convolved with the best-fit $P_\mathrm{orb}$, are shown in the left panel of Fig.~\ref{fig:orbital_profile}. Color points marked with ``O~$N$'' stand for corresponding fluxes observed during orbit $N$ ($T_0 =$ MJD~$ 54587.0$ as in~\citealt{bongiorno11}). The flux orbital profile from the binned average over all data points is shown as green points with horizontal error bars on both panels of Fig.~\ref{fig:orbital_profile}. 

For the measured value of $P_\mathrm{orb}$, two observations with \cha and \xmm appear to have been taken during the first narrow X-ray maximum (C1) and close to a deep minimum (X1). In the following we will refer to these observations as high- and low-state X-ray observations of \hessj, respectively.

\subsection{$n_H$ and spectral slope orbital profiles}
Besides obtaining flux orbital profiles, the \swift-XRT data allow us to trace the orbital variation of the spectral slope $\Gamma$ and of the hydrogen column density \nh. We split the orbit of \hessj into 10 bins and grouped the \swift observations accordingly. For each orbital bin we fitted the observations with an absorbed power-law model fixing the slopes and hydrogen column densities to be the same for all observations in one bin, while the flux levels were allowed to vary between observations. The obtained orbital profiles for \nh and $\Gamma$ are shown in Fig~\ref{fig:orbital_nh_idx} with green points. Black points show the high-quality results from individual observations of \xmm, \cha, and \suz. These high-quality data are in a good agreement with profiles measured by \swift.

The thin solid blue lines illustrate the best fit of the corresponding orbital profile with a constant, respectively (derived from only taking \swift data into account). Constant orbital profiles can be excluded at a significance level of $\sim 4.9\sigma$ (\nh orbital profile) and $\sim 4.2\sigma$ (spectral slope orbital profile). 

\section{Discussion}
\label{sec:discussion}
\subsection{``inclined disk'' model description}
The observed keV-TeV emission of \hessj was historically first modeled by~\citet{moritani15} in terms of the flip-flop scenario. In this scenario, the compact object is a neutron star, acting as a non-accreting pulsar far from the companion star. Close to periastron, at phase $\phi_0\sim 0$ according to ephemeris of~\citet{casares12}, the strong gas pressure overcomes the pulsar-wind ram pressure, quenches the pulsar wind, and suppresses the high-energy emission.
The second minimum corresponds to the apastron passage, where the energy densities of both the magnetic and soft photons fields are low. 

We suggest here an alternative interpretation, adopting the model proposed by \citet{chernyakova15} for \psrbl. 
The two-peaked orbital X-ray light-curve would be explained by the inclination of the orbital plane of the compact object relative to the Be star's disk. In this case, when the compact object's orbit crosses the disk, the higher ambient density leads to enhanced particle acceleration via wind-wind interaction. In this model, the X-ray/TeV peaks in the light-curves correspond to the first and second crossings of the disk.

According to the ``inclined disk'' model, the positions of the light-curve peaks constrain the position of the periastron. Because of the higher orbital velocities of the compact object at phases close to the periastron, the latter has to be located at phases 0.4 -- 0.5, i.e.\, at the shortest orbital separation between the peaks. The relative widths of the peaks constrain the eccentricity of the orbit to values of around $\sim 0.5$, see Fig.~\ref{fig:orbit} for a sketch of the orbital geometry. These parameters differ significantly from the ephemerids of \citet{casares12}, but are in reasonable agreement with the orbital parameters obtained from the very recent optical monitoring of the system by \citet{moritani17} for the reported here orbital period.

\subsection{Be star disk structure}

For a system orientation close to edge-on, (like the inclination of $i\gtrsim 47^\circ$ with preferred values of $i\sim 71^\circ-90^\circ$ as reported by \citealt{casares12}), the hint for a second peak observed in the \nh orbital profile is also readily understood. To illustrate this we have considered a simple geometrical model similar to that used by \cite{we_lsi} for LS~I+61\,303. In this model, at each orbital phase the observed \nh value is given by the integration of the inclined Be star disk density profile along the line of sight to the observer. We assume that the disk consists of non-ionized hydrogen and has an exponential density profile typical of an isothermal atmosphere:
\begin{equation}
    n(r,z) = n_0 \exp(-r/r_0 - |z|/z_0)
    \label{eq:disk}
\end{equation}
The characteristic radius and thickness of the disk was chosen to be $r_0 = 4\cdot 10^{13}$~cm and $z_0 = 0.4\cdot 10^{13}$~cm, and the central density as $n_0=0.2\cdot 10^{8}$~cm$^{-3}$, close to values typical for other Be stars~\citep{rivinius13}. For these parameters, the effective opening angle of the disk can be estimated as $\sim 2z_0/r_0 \sim  10^\circ$, in agreement with estimates for other Be stars~\citep{hanuschik96}. The angle between the line of the disk--orbit plane intersection and major axis of the orbit was taken to be $30^\circ$. The plane of the disk was assumed to be perpendicular to the orbital plane for simplicity. The observer was assumed to be at $i=90^\circ$ with the line of sight coinciding with the disk-orbit intersection line.

The density of the disk at the phase  corresponding to the first X-ray maximum is $n_\mathrm{max}\sim 10^{7}$~cm$^{-3}$.  For such a simple model, the sum of constant Galactic \nh (assumed to be $0.2\cdot 10^{22}$~cm$^{-2}$) and locally variable \nh is shown with a dashed blue line in the left panel of Fig.~\ref{fig:orbital_nh_idx}. %%%%%%%%%%%%%%%%%%%%%%%%%%%%%%%%%%%%%
\begin{figure}
\includegraphics[width=1.05\linewidth]{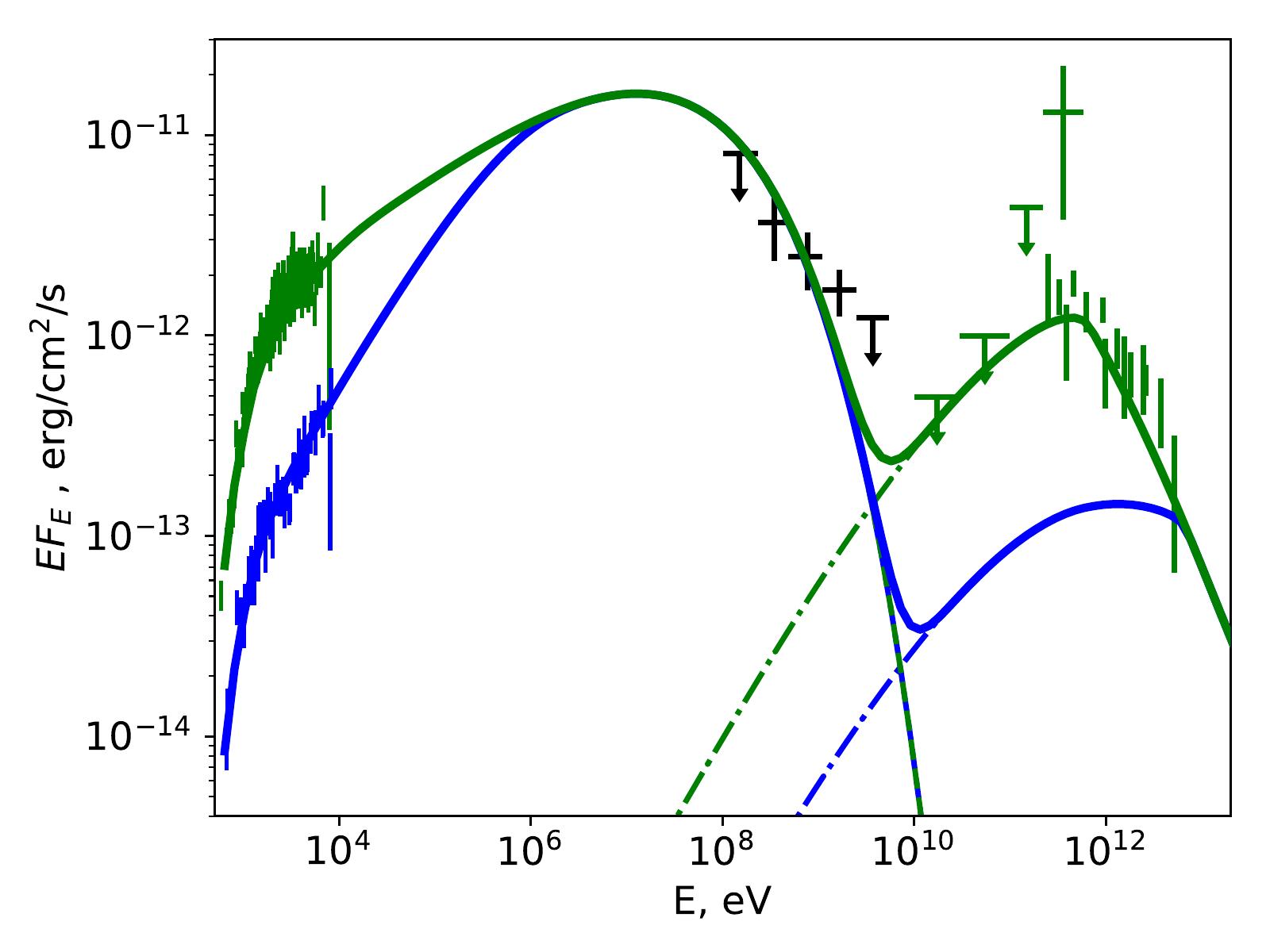}
\caption{X-ray to TeV spectral energy distribution of \hessj during its high (green points) and low (blue points) X-ray states. The X-ray data are from the \cha (C1) and \xmm (X1) observations analyzed in this work, 
the GeV data is adopted from \citet{li17} (mean spectrum, black points) and~\citet{we_hessj} (green upper limits). TeV data are adopted from \citet{veritas15}. The solid lines show the expected ``inclined disk'' model flux for the two states, while the dashed and dot-dashed lines illustrate the contributions from the synchrotron and IC model components, respectively, see Sect.~\ref{sec:discussion} for details.}
\label{fig:model}
\end{figure}
%%%%%%%%%%%%%%%%%%%%%%%%%%%%%%%%%%%%%

According to this absorption model, a double-peaked \nh profile is not expected for non-inclined disks with monotonic density profiles, as those usually assumed in the ``flip-flop'' model. Yet, non-monotonic complex dynamic density distributions can be formed by strong disk-compact object outflow interactions \citep[as seen in hydrodynamic simulations, e.g.][]{valenti17} and this could be qualitatively invoked to explain the double-peaked \nh profile. %However, strong compact object outflows have been suggested at all orbital phases by \citet{valenti17}, whereas in the ``flip-flop'' model this outflow is expected to be minimal close to periastron. 
Detailed numerical simulations would be needed to explain the observed double-peak $n_H$ density profile in the framework of the ``flip-flop'' model. 

One should also note that orbital solution with periastron at $\phi\sim 0.4$ is still in agreement with the ``flip-flop'' model. Indeed, in this case the model holds with the assumption that the gas-quenched pulsar flux at periastron is lower than the flux in the low-density apastron region.

The disk density at the first X-ray maximum, $n_{max}\sim 10^7$~cm$^{-3}$ obtained from the simple estimate above allows us to estimate the maximum spin-down luminosity of the neutron star still consistent with the accretion phase according to the flip-flop model. Following~\citet{valenti11} we estimated this luminosity assuming momenta equality of the neutron star wind and Be star outflows at the gravitational-capture radius~\citep{bondi44}:  
\begin{equation}
    %L_{sd} \lesssim 4\pi\left(\frac{2GM}{v^2}\right)^2(\rho_{max})v^2 \approx 5\cdot  
    L_\mathrm{sd}\lesssim 5\cdot 10^{26}\left(\frac{n_{max}}{10^{7}\mbox{cm}^{-3}}\right)\left(\frac{v}{c}\right)^{-2}\left(\frac{M}{M_\odot}\right)^2 \mbox{erg/s}
\end{equation}
where $v$ is the velocity of the Be star outflow and $M$ is the mass of the compact object. Assuming $v\sim 10^{-3}c$, typical for Be stars disk outflows~\citep{puls09}, and $M\sim 3M_\odot$~\citep{moritani17} this leads to a maximum spin-down luminosity allowing accretion of $L_\mathrm{sd}\lesssim 5\cdot 10^{33}$~erg/s.

The maximum of the SED of \hessj is likely located in the GeV band, see ~\citet{li17} and Fig.~\ref{fig:model}. This allows an estimate of the spin-down luminosity of the compact object as $L_\mathrm{obs}\sim 3\cdot 10^{33} (D/2.4 \mbox{kpc})^2 \mbox{erg/s}$. The luminosity estimated from the observation is hence comparable to the maximum spin-down luminosity which still allows accretion. However, given the uncertainties of the Be star disk's density, of the outflow velocity and of the distance to the binary system, we cannot conclude that the inclined disk model is definitively ruled out by this estimation.

Thus, the orbital $n_H$ profile data and the orbital solution are not sufficient to distinguish between the previously described emission models. Below we present SED-based arguments which can help to discriminate between these models.

%%%%%%%%%%%%%%%%%%%%%%%%%%%%%%%%%%%%%
\begin{figure}
\includegraphics[width=1.05\linewidth]{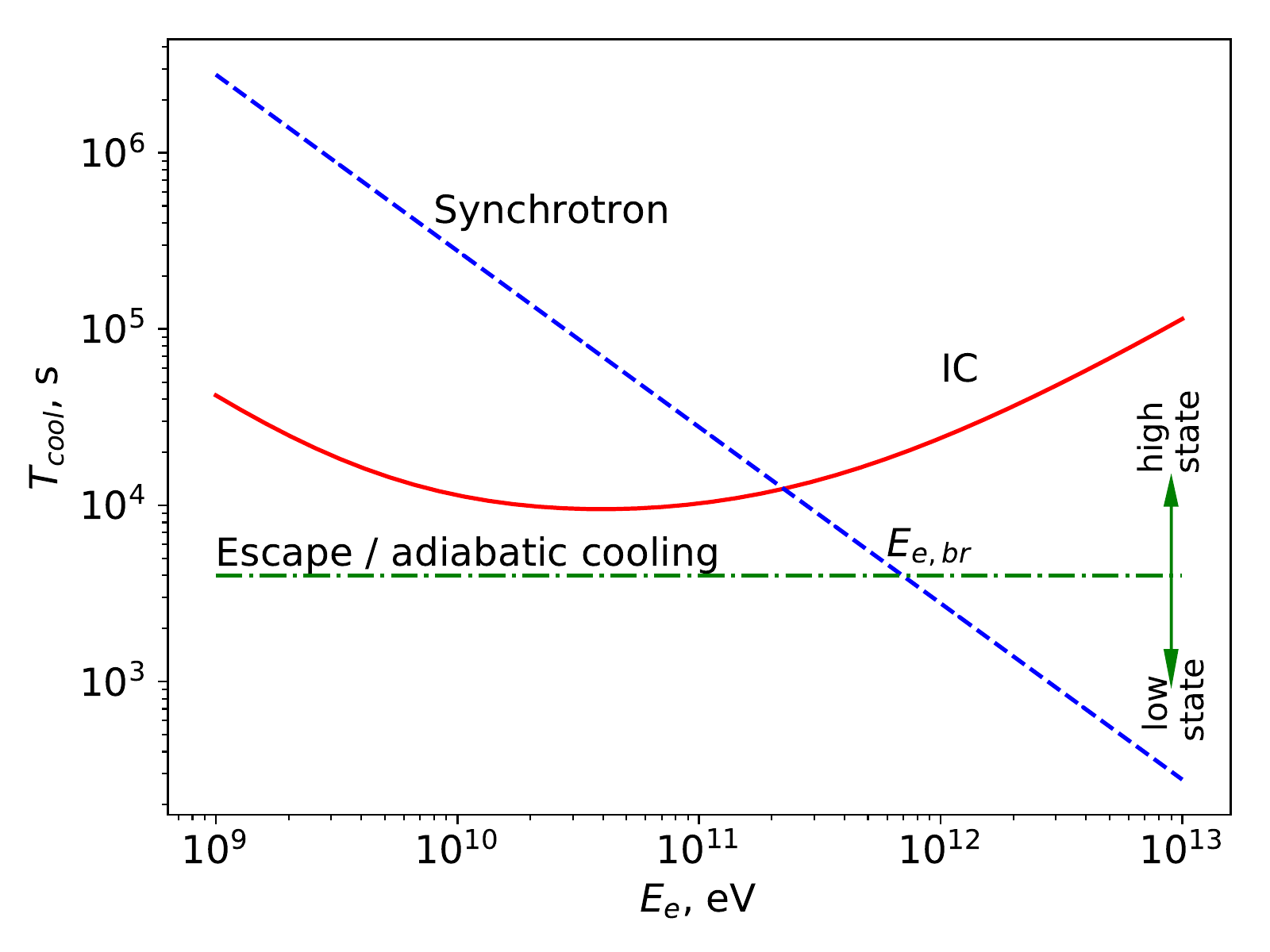}
\caption{Synchrotron, IC, and escape/adiabatic cooling times for the system parameters considered to fit the X-ray to TeV data within the ``inclined disk'' model. The vertical arrows illustrate the increase/decrease of the escape time for the high (inside the Be star disk) and low (outside the disk) states which lead to a consequent shift of the break energy $E_{e,br}$ in the cooled spectrum of electrons. }
\label{fig:tcool}
\end{figure}
%%%%%%%%%%%%%%%%%%%%%%%%%%%%%%%%%%%%%

\subsection{Spectral modelling}
The keV-TeV SED of \hessj in its high and low states is shown in Fig.~\ref{fig:model}. The X-ray data correspond to observations C1 (green points, high state) and X1 (blue points, low state). The TeV~\citep{veritas15} and $>10$~GeV data~\citep{we_hessj} are from the phase intervals 0.2--0.4 and 0.6--0.8. They are to a sufficient degree of accuracy representative of the high state. The GeV data in the 100\,MeV--10\,GeV range \citep{li17} are an average over all phase bins.

The GeV-TeV shape of the spectrum indicates the presence of a break or a cut--off located at $E_\mathrm{br} \sim 140-200$~GeV with a lower-energy spectral slope $<1.6$~\citep{we_hessj}. 
The presence of such a break might reflect a break at $E_\mathrm{e,br}$ in the spectrum of the emitting particles arising from the energy dependence of cooling losses. Due to the orbital modulation of the environmental conditions, we expect that losses' strength and thus the break energy vary along the orbit. These variations can be used to discriminate between the ``flip-flop'' and ``inclined disk'' models.

In Fig.~\ref{fig:model}~we show model spectra expected from a broken cutoff power-law distribution of electrons, calculated with the \texttt{naima v.0.8} package \citep{naima}, which uses cross-sections and SED analytic approximations for the IC and synchrotron emission by~\citet{aharonian81,aharonian10,khangulian14}. 

X-ray/GeV and GeV/TeV branches of the spectrum are produced via synchrotron and inverse Compton mechanisms, respectively, and are shown in Fig.~\ref{fig:model} with dashed and dot-dashed lines. For simplicity, the magnetic field and soft photon density is assumed to be the same along the orbit and the inverse Compton scattering is assumed to be isotropic.

In general, anisotropic IC effects can be quite important for close to edge-on system orientations, see e.g.~\citet{khangulian08,dubus08a}. The quality of the available TeV data (no statistically significant difference of the \hessj spectra for the first and second X-ray maxima; no detection of \hessj in the low state) does, however, not allow effects of anisotropic IC or of variations of the soft photon SED along the orbit to be studied in detail.

The very high energy and \cha observations of \hessj in the high state fix the high energy slope of the electron spectrum to $\Gamma_\mathrm{e,2}\approx 2.3$, the position of the break to $E_\mathrm{e,br}\approx 0.7$~TeV (for IC scattering on photons of the Be star with a temperature T$\sim 3\cdot 10^4$~K), and the high-energy cutoff to $E_\mathrm{cut}\sim 100$~TeV. The magnetic field strength in the emitting region can be assumed to be $B\approx 0.3$~G. To match the low state X-ray data, in this state the break has to be located at an order of magnitude higher energies~$\gtrsim 7$~TeV.  In the calculation we assumed the emitting region to be located at a distance $d\sim 2.5$~a.u. from the Be star which has an assumed typical luminosity of $L_*=3\cdot 10^{38}$~erg/s. 

\subsection{Origin of the break in the electron spectrum}
\label{sec:break_origin}
Fixing the low-energy slope of the electron distribution to that matching the \xmm low-state spectrum, namely to $\Gamma_\mathrm{1,e}\approx 1.3$, we can robustly meet the constrains given by the \fermi-LAT upper limits in the 100~GeV range. The softening of the electron spectrum by $\Delta\Gamma\simeq 1$ in the high state compared to the low state is typical of synchrotron cooling \citep{Blumenthal}. A similar softening around periastron has also been observed in the \psrbl system \citep{chernyakova15}. Synchrotron cooling modifies the electron spectrum above $E_\mathrm{e,br}\gtrsim 1$~TeV, but leaves the spectrum below 1\,TeV unaffected.
The absence of cooling in the energy band below 1~TeV could be attributed to the escape of sub-TeV electrons from the system. 

The synchrotron and IC cooling times (using the approximation of~\citealt{khangulian14}), as well as the energy-independent escape timescale are shown in Fig.~\ref{fig:tcool} for the aforementioned high-state model parameters. For sub-TeV electrons, the escape timescale is much shorter than synchrotron/IC cooling timescales and the spectrum does not suffer from corresponding cooling losses. At energies higher than the break energies $E_\mathrm{e} > E_\mathrm{e,br}$, the synchrotron cooling time is shorter than the escape time, which leads to a significant softening of the spectrum. The vertical arrows illustrate the change of the escape time as the source transits from high state (the compact source is in the Be star disk; longest escape time) to the low state (outside of the disk; shortest escape time). These changes in the escape time lead to a consequent shift of the break energy $E_\mathrm{e,br}$ to lower or higher energies, respectively.

Alternatively, a spectral break in the electron spectrum can occur at energies at which energy losses changes from the adiabatic to the synchrotron cooling regime, see Fig.~\ref{fig:tcool} and e.g.~\cite{khan07} and \cite{Suzaku2009} for the cases of \psrbl and LS~5039, respectively. The adiabatic loss time can be shortest in the sparse regions outside of the Be star disk and longest in the dense regions inside the disk. This would also explain the shift of the spectral energy break to higher energies in the source's low state.

We note that an IC/synchrotron cooling times transition for the considered model parameters is expected at $\gtrsim 0.1$~TeV energies. Considering this as an order-of-magnitude estimate it is possible to associate the break in the electron spectrum with this transition. An increased soft photon density at phases close to periastron (coinciding with the low state) can lead to a decrease of the IC cooling time. The periods of crossing of the optically-thick disk by the compact object can correspond to the local maxima of IC cooling at certain orbital phases.
We would like to note thus that the orbital changes in IC/synchrotron cooling times transition is very similar to the described above alternatives. Similar transition effect was studied in the context of other gamma-ray binary systems in detail in~\citet{khangulian05,khangulian08}.

\subsection{Orbital variability of the SED}

For all break energy interpretations described in Sec.~\ref{sec:break_origin}, we expect that when the compact object enters the denser regions of the disk (i.e.\ at phases $\phi\sim 0.3$ and $\phi\sim 0.6$) the break energy would be located between its minimum and maximum values observed in the high (deep in the disk) and the low (out of the disk) states, and could be detected as a break in the X-ray photon spectrum in other observations. We find that \suz data taken at phase $\sim 0.3$ (S2 observation) marginally prefer ($\Delta cstat=5$ for 2 d.o.f.) the model with a break over a single power-law model, with an indication of a break at $\sim 5$~keV. 

\nus observations N1 and N2 taken at phases $\sim 0.15$ and $\sim 0.25$ exhibit significantly softer slopes than the ones observed with \swift and \suz at the same phases. The energy range of \nus is $\sim 2-50$~keV, significantly above that of the other instruments. These observations are therefore representative of energies above the break. \nus data in fact constrain the position of the break (assuming a $\Delta\Gamma < 0.4$) to $\lesssim 6$~keV. Joint \xmm and \nus observation of the source are necessary to firmly detect or reject the presence of the spectral break at orbital phases between the high and low states.

The deviation from a power-law shape of the spectrum may also systematically affect the spectral slopes measured with \swift. This may explain some of the discrepancy in the slope measurements between the mean over several orbits with \swift and the individual measurement (C1) of the slope with \cha at phase $\sim 0.4$, see Fig.~\ref{fig:orbital_nh_idx}.
The observed discrepancy may also be explained either by strong gradients in the orbital profile of the spectral slope, or by a systematic variation of the spectral slope from orbit to orbit. 
A broken power-law shape of the spectrum can be expected for both ``flip-flop'' and ``inclined disk'' models, since interpretations of the break origin are valid for both models. However, the two models can be distinguished by observations of the spectral break variability along the orbit.

In the ``flip-flop'' model it is expected that the escape/adiabatic cooling time from the system increases as the compact object enters the dense disk regions and approaches periastron, before the accretion starts. This leads to a gradual shift of the X-ray/TeV spectral break energy to lower values as the distance between the compact object and the star decreases. For the poor-statistics data the shift of the break can manifest itself by the gradual softening of the spectrum. The slope orbital profile monotonically softens and the break energy shifts to lower energies from apastron to periastron.

On the contrary, in the ``inclined disk'' model the periastron is characterized by low density values and, therefore, by relatively short escape/adiabatic cooling times. The longest escape times (lowest $E_\mathrm{e,br}$) are expected during the compact object's disk crossing, i.e. at phases corresponding to the maxima of the X-ray/TeV light-curves of the system. Observationally, the X-ray/TeV spectral break energy exhibits a non-monotonic orbital profile with two local minima (corresponding to disk crossing or maxima of X-ray lightcurve) and two maxima (at periastron and apastron phases). Similar non-monotonic behavior is expected in this case for the spectral slope in a low-statistics case.
The double-peak shape of the $n_H$ orbital profile, if confirmed, can also support the ``inclined disk'' model.

To conclude, detailed orbital profile observations of the position of $E_\mathrm{e,br}$ would permit us to distinguish between the two presented models. We would like to stress, that these observations can be equally performed either in the TeV or in the X-ray energy band, by observing the corresponding feature in either the synchrotron or in the IC branch of the SED.
\newline
\newline
\textit{Acknowledgements}. This work was partially supported by the EU COST Action (COST-STSM-CA16104-38088) ``GWverse''. This research has made use of the XRT Data Analysis Software (XRTDAS) developed under the responsibility of the ASI Science Data Center (ASDC), Italy. This work was supported by the Carl-Zeiss Stiftung through the grant ``Hochsensitive Nachweistechnik zur Erforschung des unsichtbaren Universums'' to the Kepler Center f{\"u}r Astro- und Teilchenphysik at the University of T{\"u}bingen. MC acknowledges the SFI/HEA Irish Centre for High-End Computing (ICHEC) for the provision of computational facilities and support. The authors acknowledge support by the state of Baden-W\"urttemberg through bwHPC and by the Eberhard Karl University of T{\"u}bingen.

% Bibliography and bibfile
\def\aj{AJ}%
          % Astronomical Journal
\def\actaa{Acta Astron.}%
          % Acta Astronomica
\def\araa{ARA\&A}%
          % Annual Review of Astron and Astrophys
\def\apj{ApJ}%
          % Astrophysical Journal
\def\apjl{ApJ}%
          % Astrophysical Journal, Letters
\def\apjs{ApJS}%
          % Astrophysical Journal, Supplement
\def\ao{Appl.~Opt.}%
          % Applied Optics
\def\apss{Ap\&SS}%
          % Astrophysics and Space Science
\def\aap{A\&A}%
          % Astronomy and Astrophysics
\def\aapr{A\&A~Rev.}%
          % Astronomy and Astrophysics Reviews
\def\aaps{A\&AS}%
          % Astronomy and Astrophysics, Supplement
\def\azh{AZh}%
          % Astronomicheskii Zhurnal
\def\baas{BAAS}%
          % Bulletin of the AAS
\def\bac{Bull. astr. Inst. Czechosl.}%
          % Bulletin of the Astronomical Institutes of Czechoslovakia
\def\caa{Chinese Astron. Astrophys.}%
          % Chinese Astronomy and Astrophysics
\def\cjaa{Chinese J. Astron. Astrophys.}%
          % Chinese Journal of Astronomy and Astrophysics
\def\icarus{Icarus}%
          % Icarus
\def\jcap{J. Cosmology Astropart. Phys.}%
          % Journal of Cosmology and Astroparticle Physics
\def\jrasc{JRASC}%
          % Journal of the RAS of Canada
\def\mnras{MNRAS}%
          % Monthly Notices of the RAS
\def\memras{MmRAS}%
          % Memoirs of the RAS
\def\na{New A}%
          % New Astronomy
\def\nar{New A Rev.}%
          % New Astronomy Review
\def\pasa{PASA}%
          % Publications of the Astron. Soc. of Australia
\def\pra{Phys.~Rev.~A}%
          % Physical Review A: General Physics
\def\prb{Phys.~Rev.~B}%
          % Physical Review B: Solid State
\def\prc{Phys.~Rev.~C}%
          % Physical Review C
\def\prd{Phys.~Rev.~D}%
          % Physical Review D
\def\pre{Phys.~Rev.~E}%
          % Physical Review E
\def\prl{Phys.~Rev.~Lett.}%
          % Physical Review Letters
\def\pasp{PASP}%
          % Publications of the ASP
\def\pasj{PASJ}%
          % Publications of the ASJ
\def\qjras{QJRAS}%
          % Quarterly Journal of the RAS
\def\rmxaa{Rev. Mexicana Astron. Astrofis.}%
          % Revista Mexicana de Astronomia y Astrofisica
\def\skytel{S\&T}%
          % Sky and Telescope
\def\solphys{Sol.~Phys.}%
          % Solar Physics
\def\sovast{Soviet~Ast.}%
          % Soviet Astronomy
\def\ssr{Space~Sci.~Rev.}%
          % Space Science Reviews
\def\zap{ZAp}%
          % Zeitschrift fuer Astrophysik
\def\nat{Nature}%
          % Nature
\def\iaucirc{IAU~Circ.}%
          % IAU Cirulars
\def\aplett{Astrophys.~Lett.}%
          % Astrophysics Letters
\def\apspr{Astrophys.~Space~Phys.~Res.}%
          % Astrophysics Space Physics Research
\def\bain{Bull.~Astron.~Inst.~Netherlands}%
          % Bulletin Astronomical Institute of the Netherlands
\def\fcp{Fund.~Cosmic~Phys.}%
          % Fundamental Cosmic Physics
\def\gca{Geochim.~Cosmochim.~Acta}%
          % Geochimica Cosmochimica Acta
\def\grl{Geophys.~Res.~Lett.}%
          % Geophysics Research Letters
\def\jcp{J.~Chem.~Phys.}%
          % Journal of Chemical Physics
\def\jgr{J.~Geophys.~Res.}%
          % Journal of Geophysics Research
\def\jqsrt{J.~Quant.~Spec.~Radiat.~Transf.}%
          % Journal of Quantitiative Spectroscopy and Radiative Trasfer
\def\memsai{Mem.~Soc.~Astron.~Italiana}%
          % Mem. Societa Astronomica Italiana
\def\nphysa{Nucl.~Phys.~A}%
          % Nuclear Physics A
\def\physrep{Phys.~Rep.}%
          % Physics Reports
\def\physscr{Phys.~Scr}%
          % Physica Scripta
\def\planss{Planet.~Space~Sci.}%
          % Planetary Space Science
\def\procspie{Proc.~SPIE}%
          % Proceedings of the SPIE
\let\astap=\aap
\let\apjlett=\apjl
\let\apjsupp=\apjs
\let\applopt=\ao
%\bibliographystyle{Wiley-ASNA}
%\bibliography{hessbib} 

\end{document}